\documentclass[aps,pre,twocolumn,notitlepage,superscriptaddress]{revtex4-1}

\usepackage{amsmath}
\usepackage{amssymb}
\usepackage{hyperref}
\usepackage{graphicx}
\usepackage[arrow, matrix, curve]{xy}
\usepackage{bbm}
\usepackage{bm}
\usepackage{yhmath}
\usepackage{color}
\usepackage{physics}
\usepackage{bbold}

\usepackage{color}

\renewcommand\vec[1]{\boldsymbol{\mathrm{#1}}}

\usepackage[usenames,dvipsnames]{xcolor}
\hypersetup{colorlinks=true, linkcolor=BrickRed, urlcolor=blue!50!black, citecolor=blue!50!black}

\usepackage[normalem]{ulem}

\makeatletter
\newcommand\hide@visible[1]{%
  \bgroup\fboxsep=.3ex\colorbox{Gray}{begin hide}%
  #1\colorbox{Gray}{end hide}\egroup%
}
\newcommand\hide@hidden[1]{%
  \bgroup\fboxsep=.3ex\colorbox{Gray}{hidden text}%
}
\newcommand\hide@invisible[1]{}
\newcommand\makevisible{\let\hide\hide@visible}
\newcommand\makehidden{\let\hide\hide@hidden}
\newcommand\makeinvisible{\let\hide\hide@invisible}
\makeatother
\makehidden

\begin{document}

\title{Run-and-Tumble Particles Learning Chemotaxis}

\author{Nicholas Tovazzi}
\affiliation{Institut f\"ur Theoretische Physik, Universit\"at Innsbruck, Technikerstra{\ss}e 21A, A-6020, Innsbruck, Austria}
\affiliation{Dipartimento di Fisica, Universit\`a degli Studi di Trento, via Sommarive 14, Trento, 38123, Italy}

\author{Gorka Mu{\~n}oz-Gil}
\affiliation{Institut f\"ur Theoretische Physik, Universit\"at Innsbruck, Technikerstra{\ss}e 21A, A-6020, Innsbruck, Austria}

\author{Michele Caraglio}
\email{Michele.Caraglio@uibk.ac.at}
\affiliation{Institut f\"ur Theoretische Physik, Universit\"at Innsbruck, Technikerstra{\ss}e 21A, A-6020, Innsbruck, Austria}

\date{\today}

\begin{abstract}
Through evolution, bacteria have developed the ability to perform chemotactic motion in order to find nourishment.
By adopting a machine learning approach, we aim to understand how this behavior arises.
We consider run-and-tumble agents able to tune the instantaneous probability of switching between the run and the tumble phase.
When such agents are navigating in an environment characterized by a concentration field pointing towards a circular target, we investigate how a chemotactic strategy may be learned starting from unbiased run-and-tumble dynamics.
We compare the learning performances of agents that sense only the instantaneous concentration with those of agents having a short-term memory that allows them to perform temporal comparisons.
While both types of learning agents develop successful target-search policies, we demonstrate that those achieved by agents endowed with temporal comparison abilities are significantly more efficient, particularly when the initial distance from the target is large.
Finally, we also show that when an additional length scale is imposed, for example by fixing the initial distance to the target, the learning agents can leverage this information to further improve their efficiency in locating the target.
\end{abstract}

\maketitle

\section{Introduction}\label{sec1}

Chemotaxis is a widespread phenomenon in nature.
Paradigmatic examples are bacteria foraging nourishment or escaping toxic substrates~\cite{Elgeti2015,Berg2004}, phagocytes of the immune system responding to injury or infection~\cite{Devreotes1988,Deoliveira2016}, and sperm cells navigating towards the egg~\cite{Eisenbach2006}.

Several bacteria, including \textit{Escherichia coli}, exhibit movement patterns characterized by an alternating sequence of near-to-straight `runs' at almost constant speed and reorientation events called `tumbles'~\cite{Berg2004}.
The durations of such phases are stochastic processes. 
For example, \textit{E. coli} in a uniform dilute aqueous medium, display exponentially distributed phase durations, with a mean of $1$ and $0.1$ seconds for the run and the tumble phase, respectively~\cite{Berg2004,Sourjik2012}.
However, in the presence of nutrients or other environmental stimuli, chemotactic motion emerges, consisting of a biased random walk with prolonged runs in the preferred direction~\cite{Berg2004,Sourjik2012,Berg1972,Block1982}.
Bacteria like \textit{E. coli} reach this goal thanks to a complex chemotaxis network that allows them to sense gradients of chemicals by making temporal comparisons on a short time scale~\cite{Sourjik2012,Segall1986,Li2011}.
Such a network is highly refined and evolutionary optimized, being able to discriminate concentrations in about five orders of magnitudes starting from about 3 ligands per cell volume~\cite{Sourjik2012}.

As a paradigmatic model of nonequilibrium dynamics, run-and-tumble motion has been extensively investigated in statistical physics, see~\cite{Cates2012} and references therein.
At the single-particle level, the typical theoretical modeling involves a diffusion-drift equation for the one-particle probability density which allows finding exact analytical results~\cite{Schnitzer1993,Tailleur2008,Tailleur2009,Malakar2018}.
Chemotactic behavior can be included by asserting that the drift velocity in that equation is proportional to the gradient of the scalar field modelling the concentration of chemoattractants~\cite{Schnitzer1993}, with a coefficient of proportionality which can be related to the microscopic parameters of the run-and-tumble particle~\cite{deGennes2004}.
The mechanism leading to such behavior is a continuous modulation of the instantaneous tumbling probability as a function of a differential weighting of past measurements of chemoattractant concentration~\cite{Schnitzer1993,deGennes2004}. 
Within this picture, the steady-state behavior and the optimal chemotactic strategy, i.e. the optimal modulation of the tumble rate in response to concentration changes, can be analytically investigated in various regimes~\cite{Schnitzer1993,Strong1998,Clark2005,Kafri2008}.
It turns out that the optimal strategy depends on properties of the environment and of the individual bacterium, and is therefore highly adaptive~\cite{Strong1998}.

In this work, to better understand how evolution shaped the search strategies of bacteria, we adopt an approach based on machine learning.
In the recent past, reinforcement learning~\cite{Sutton2018} and genetic algorithms~\cite{Mitchell1998} have already emerged as powerful tools in active matter research~\cite{Cichos2020,Tsang2020review}.
Focusing on improving the navigation and target-search performances of microswimmers, promising results have been obtained in several situations including homogeneous environments with unknown target position~\cite{munozgil2023,Kaur2023,Caraglio2024}, simple energy landscapes~\cite{Schneider2019}, viscous surroundings~\cite{Muinos-Landin2021,Tsang2020,Hartl2021}, steady or turbulent flows~\cite{Colabrese2017,Gustavsson2017,Colabrese2018,Biferale2019,Alageshan2020}, and complex motility fields~\cite{Monderkamp2022}.
Interestingly, in a viscous environment involving chemical gradients, biased run-and-tumble dynamics emerges when applying a genetic algorithm to a simple three-bead swimmer able to perform one-dimensional locomotion by adapting the length of the arms connecting the central bead to the external ones~\cite{Hartl2021}.

Here, we consider run-and-tumble agents modelled as intermittent active Brownian particles~\cite{Caraglio2024,Kiechl2024} and initially performing unbiased random dynamics.
Exploiting the Projective Simulation (PS) algorithm~\cite{Briegel2012}, we show how an efficient chemotactic target-search behavior can be achieved when the agents are immersed in an environment characterized by a single circular target releasing chemicals uniformly into the surrounding space.
To better understand the role of temporal comparison, we benchmark the performance of agents that sense only the instantaneous concentration against those equipped with short-term memory, enabling them to perform temporal comparisons.

\section{Model}\label{sec2}

A circular target of diameter $\sigma$ is placed at the origin of an infinite two-dimensional domain in which the agent is moving.
With two-dimensional Run-and-Tumble motion in mind, we model our agent as a particle that alternates between two distinct phases encoded in the binary variable $\phi$: 
A passive phase ($\phi = 0$), during which the motion of the particle is governed by standard translational diffusion with diffusion coefficient $D$, and an active phase ($\phi = 1$), during which the particle is also able to self-propel with constant velocity $v$ in a given direction described by an angle $\vartheta$.
The latter also undergoes a diffusion process with rotational diffusion coefficient $D_{\vartheta}$. 
These two navigation modes are also referred to as the Brownian Particle (BP) phase and the ABP phase, respectively.
When the particle changes phase, a complete randomization of the self-propultion direction occurs (tumble event).
We impose that target detection is enabled only when the agent is in the BP phase.
In contrast, during the ABP phase, the agent cannot acquire the target but on the other hand, exploiting its self-propulsion, it can rapidly relocate to a different region in space.
This choice places our model within the class of intermittent-search strategies~\cite{Benichou2005,Benichou2006,Benichou2007,Benichou2011}, which are based on the idea that fast motion, while allowing for quick relocation, degrades perception abilities.
 
We endow our run-and-tumble particles with some perception abilities and investigate implications on their target-search behavior.
In particular, we consider two sets of agents (A and B) with different sensorial skills.
Agents of type A are able to sense the distance from the target, $r$, which is a proxy for the concentration of chemicals released by nourishment.
At each time step $t$, their \textit{state} $s^{\rm A}_t$, is characterized by the tuple $s^{\rm A}_t=(\phi_t, r_t)$, where $\phi_t$ represents the current phase and $r_t$ represents the distance from the target.
Agents of type B have the additional ability to sense concentration differences on a short time scale, which is attained by evaluating if, after an integration time step $\Delta t$, the agent reaches a position that is closer to the target.
Their state $s^{\rm B}_t$, is given by the tuple $s^{\rm B}_t=(\phi_t, r_t, \omega_t)$, with $\omega_t$ a binary variable that equals one if the agent moved closer to the target ($r_t<r_{t-\Delta t}$) and zero otherwise.
Agent of type B are thus mimicking the perception abilities of the bacteria that, like the \textit{E. coli}, are able to make temporal comparison of concentrations.
Following the reinforcement learning (RL) framework~\cite{Sutton2018}, given its current state $s_t$, the agent responds with an \textit{action} $a_t$, receiving a \textit{reward} if this action proves beneficial.
Here, the action involves deciding whether to retain the current phase or switch to the other one, thus also randomizing the self-propulsion direction.
This decision follows a probabilistic rule, with $p_t$ the probability of switching phases.
Note that $p_t$ is not fixed: It depends on the agent's current state, $s_t$. 
The entire set of these state-dependent probabilities constitutes the agent's \textit{policy}, which is updated during learning to maximize the total reward (see Methods section for details).

Integrating these concepts into the standard ABP model~\cite{Bechinger2016} within a homogeneous environment yields the following Langevin equations, discretized according to the It\^{o} rule: 
\begin{eqnarray}\label{eom1}
	\phi_{t+\Delta t} & = & \left\lbrace 
	\begin{array}{ll}
		\phi_t & \mbox{with probability } 1-p_t \; , \\
		1-\phi_t & \mbox{with probability } p_t \; ,
	\end{array}
	\right. \\
	\label{eom2}
	\vec{r}_{t\!+\!\Delta t} &=& \vec{r}_{t} + v \, \vec{u}_{t} \, \phi_{t} \, \Delta t  + \sqrt{2D\Delta t} \, \boldsymbol{\xi}_t \; , \\ 
	\label{eom3}
	\vartheta_{t\!+\!\Delta t} &=& 
	\left\lbrace 
	\begin{array}{ll}
		\vartheta_{t} + \sqrt{2D_{\vartheta}\Delta t} \, \eta_t  & \mbox{if } \phi_{t+\Delta t} = \phi_t  \; ,\\
		2\pi \zeta_t & \mbox{otherwise} \; .
	\end{array}
	\right. 
\end{eqnarray}
Here, $\vec{r}_t = (x_t,y_t)$ is the position at time $t$ and $\vec{u}_{t} = \big(\cos\vartheta_{t},\sin\vartheta_{t}\big)$.
Additionally, the components of the vector noise $\boldsymbol{\xi}_{t}=(\xi_{x,t},\xi_{y,t})$ and of the scalar noise $\eta_t$ are independent Gaussian random variables with zero mean and unit variance. 
Finally, $\zeta_t$ is a random variable distributed according to a uniform distribution in the interval $[0, 1]$.
Note that when the particle is in the passive Brownian phase ($\phi_t=0$), spatial evolution is decoupled from the orientational diffusion of the self-propulsion.
In the following, we set the unit of length as the target size $\sigma$ and the unit of time as $\tau := \sigma^2/4D$, the typical time a passive particle requires to traverse this distance.
The model has then two dimensionless parameters: the P\'eclet number $\text{Pe} := v\tau/\sigma$, representing the importance of active relocation with respect to passive diffusion during an ABP phase, and the persistence $\ell^* := v/D_{\vartheta}\sigma$, reflecting the persistence of directed motion in the ABP phase.

The learning process is split into several \textit{episodes}, each episode starting with the agent placed at a certain position $\vec{r}_{\rm ini}$, uniformly sampled in the annular region defined by $\sigma/2 < r_{\rm ini} \leq \tilde{R}$, and having a certain self-propulsion direction $\vartheta_{\rm ini}$, uniformly sampled in the interval $(0,2\pi]$.
Here, $\tilde{R}=10\sigma$ represents the maximum initial distance from the target that has been considered.
We also use this parameter to define the maximum distance beyond which the agent can no longer detect the distance to the target.
Beyond this distance, for each $\phi$ (and $\omega$), all the values of $r$ are collected in a single state.
Each episode ends either when the agent meets the target ($r \leq \sigma/2$) or, in any case, after a time $\tilde{\tau} = \tau$, which, on average, is long enough to allow our run-and-tumble particles to travel a distance larger than $\tilde{R}$ provided that $\text{Pe} \gtrsim 10$ and that active and passive phases have a typical duration smaller than $\tilde{\tau}$.
Each time the agent finds the target it earns a positive reward.
Since rewards are given only when the target is found, the agent learns to minimize the search time and, in effect, to optimize search efficiency.

\begin{figure*}[h!]
\centering
\includegraphics[width=0.9\textwidth]{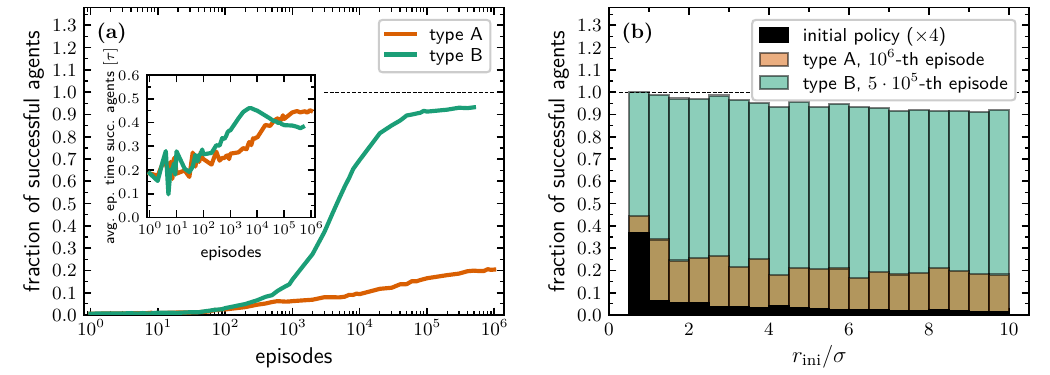}
\caption{
{\bf a)} Fraction of successful agents as a function of the number of episodes for both type A (orange line) and type B (green line) agents. 
Inset: Average episode time for successful agents as a function of the number of episodes.
{\bf b)} Fraction of successful agents as a function of initial distance from the target, $r_{\rm ini}$, evaluated considering the initial policy (black bars) and the one learned at episode $10^6$ for type A (orange bars) and at episode $5 \cdot 10^5$ for type B (green bars) agents. To improve visibility, values related to the initial policy are multiplied by a factor $4$.
}\label{fig1}
\end{figure*}

\section{Results}\label{sec3}

In the following, the reported results are evaluated by averaging the different quantities over a sample of $N=10^4$ independent learning agents with $\text{Pe}=100$ and $\ell^*=1$.
The latter parameters are such that, similar to real bacteria, our agents are alternating phases of standard diffusion and runs during which the direction of the motion slowly varies.
The details of the learning algorithm PS are reported in the Methods section.
Here, we just note that the initial policy is chosen so that the ratio of time spent in the active phase to that in the passive phase matches the ratio of time \textit{E. coli} bacteria spend in the run phase versus the tumble phase.

We start by considering how the fraction of agents ending an episode with target acquisition evolves during the learning process.
Both for agents of type A and for agents of type B we observe a steady increase in the performances, with agents able to sense their phase and the distance to the target (type A) showing an improvement from about $0.6$\%, corresponding to the adopted initial policy, to about $20$\% at the $10^6$-th episode, see Figure~\ref{fig1}a.
Thanks to their additional perceptor, making the agents aware if over a time step $\Delta t$ they are getting closer or not to the target, agents of type B display an impressive learned efficiency with about $94$\% of found targets at the $5 \cdot 10^5$-th episode.
The performances of type B agents are even more remarkable if one considers that, in the active phase with $\text{Pe}=100$, over a single time step $\Delta t$ the distance covered due to the self-propulsion mechanism is equal to the average distance covered by translational diffusion. 
Thus, being the latter in a completely random direction, the exploitable information transmitted to the additional perceptor in the form of the binary variable $\omega$ is greatly reduced.
Note that while the type B agents display a plateau of efficiency after about $10^5$ episodes, type A agents are still improving their performances at episode $10^6$. 
However, due to the exponential slowing down of the learning efficiency, it becomes computationally impractical to check when type A agents plateau by extending the number of episodes by another one or two orders of magnitude.

While the number of successful episodes increases, the average time to reach the target in successful episodes is initially also increasing, and, only for type B agents, after about $600$ episodes it starts to decrease, see inset of Figure~\ref{fig1}a.
This result, at first counterintuitive, is due to the fact that at the beginning of the learning process, finding the target is mainly a matter of favorable initial conditions, i.e. a value of $r_{\rm ini}$ slightly larger than the target radius $\sigma/2$, see Figure~\ref{fig1}b.
On the other hand, while learning, more and more agents having a larger initial distance from the target become successful (Figure~\ref{fig1}b), which initially results in an increase of the average time to reach the target.
However, at large episode numbers, type B agents become more and more able to exploit the active phase in order to quickly cover the distance separating them from the target, with the consequent decrease in the average time to reach the target.
Interestingly, while for particles of type A the final fraction of successful events shows a clear peak for small initial distances and a rather flat distribution for other distances, the same observable computed for type B agents remains close to one and only very slowly decreases with increasing distance, see Figure~\ref{fig1}b.

\begin{figure*}[h!]
\centering
\includegraphics[width=0.9\textwidth]{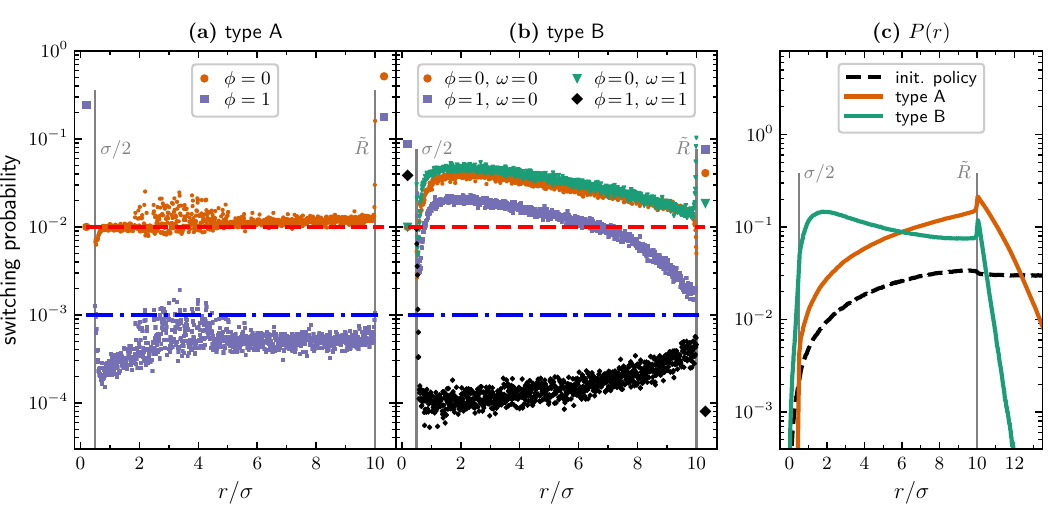}
\caption{
{\bf a)} Switching probabilities from passive to active phase (orange circles) and from active to passive phase (blue squares) for type A agents at episode $10^6$.
{\bf b)} For type B agents at episode $5\cdot10^5$, switching probabilities from passive to active phase with $\omega=0$ (orange dots) and with $\omega=1$ (green triangles), and from active to passive phase with $\omega=0$ (blue dots) and with $\omega=1$ (black diamonds).
In both panel (a) and (b), we also report the initial probability of switching from the passive to the active phase (dashed red line) and from the active to the passive phase (dash-dotted blue line).
The value of the switching probability inside the target ($r\leq\sigma/2$) and for $r>\tilde{R}$ are reported with a larger symbol.
{\bf c)} Radial distribution of type A and type B agents during target search events.
Statistics obtained over $10^5$ target-search events adopting the policies reported in panels (a) and (b). Normalization is such that $ \int_0^{\infty} \text{d}r  \, P(r) = 1 $.
}\label{fig2}
\end{figure*}

To gain a deeper understanding of the behavior of the trained agents, we can leverage the interpretability nature of the PS scheme and directly examine the learned policy, i.e. the probabilities of switching phase given their state.
These are computed starting from the matrix encoding the learning process (the $H$-matrix, see Methods section for details), averaged over different agents, and are reported in Figure~\ref{fig2}a and~b for type A and type B agents, respectively.

Starting with type A agents (Figure~\ref{fig2}a), we can observe that the switching probability inside the target area is very high when the agent is in the active phase ($\phi=1$) and is equal to the relative value of the initial policy when the agent is in the passive phase ($\phi=0$).
Intuitively, these observations are respectively in line with the facts that the agent cannot detect the target while being in the active phase, thus willing to become passive as soon as it enters the target area, and that when the agent finds the target the episode ends and the H-matrix value corresponding to $\phi=0$ and $r\leq \sigma/2$ is thus never updated.
We also note that the switching probability is very high in both phases when $r>\tilde{R}$, meaning that, when its distance from the target becomes too large, the agent tumbles frequently, attempting to find a favorable run direction that brings it back to the region where it is able to better resolve the distance from the target.
In the region defined by $\sigma/2 < r \leq \tilde{R}$, we can at first notice that the switching probability when being in the passive phase remains close to its initial value while the switching probability when being in the active phase is generally smaller than its corresponding initial one, meaning that active phases have to be on average longer than what initially defined.
Furthermore, we see a quite large volatility in the values of both the switching probabilities, especially for intermediate distances between $2\sigma$ and $5\sigma$.
This observation is likely due to the fact that different agents learn quite different policies in terms of the specific value of the switching probability corresponding to the various distance bins and that this value for $2\sigma \lesssim r \lesssim 5\sigma$ is affecting relatively less the agent performances.
A more careful inspection shows a small drop of the switching probability for $\phi=0$ when approaching $r=\sigma/2$ and a corresponding increase of the one for $\phi=1$.
This results in the agents willing to have a relatively more passive behavior (which allows target detection) in the close vicinity of the target.
Finally, we notice a small increase in both the probabilities, resulting in relatively more frequent tumbles, just before reaching the distance $r=\tilde{R}$.

When considering the policy learned by type B agents (Figure~\ref{fig2}b) we can detect similar features to those previously detailed for type A agents.
Namely, a large probability to switch from active to passive inside the target area, increased frequency of tumbling events in the outer region, and a relatively large spread of the probabilities values in between these two extrema.
However, the introduction of the additional perceptor comes with new features not displayed by the first kind of agents.
Firstly, although there is no immediately obvious distinction between the probability of switching from the BP to the ABP phase when the particle is getting closer to the target ($\omega=1$) or not ($\omega=0$), type B agents further reduce the duration of the passive phase. 
This enables them to spend more time in the active phase, where they can take advantage of the additional information provided by the extra perceptor.
Indeed, the probability of switching from the ABP to the BP phase is clearly distinct if $\omega=0$ or $\omega=1$.
Agents that are active and are getting closer to the target decrease their switching probability, such that they can exploit their self-propulsion to decrease even more their distance from the target.
In contrast, agents that are active but are not getting closer to the target increase the switching probability with the aim of tumbling and starting as soon as possible a new run eventually having a more favorable direction.

Finally, we note that the probability of switching from the BP to the ABP phase, just before $r=\tilde{R}$, shows a drop if $\omega=0$ and an increase otherwise.
This behavior is not easy to interpret because one would expect that, being each step in the passive phase in a random direction, there should be no difference in the switching probabilities shown for $\omega=0$ and for $\omega=1$.
However, the variable $\omega$ is evaluated as a temporal comparison between two different time integration steps and there is a certain probability that the phase at the two times differs.
Furthermore, in PS the rewards are propagated back in time through the glow matrix (see Methods section for details).
Thus,  favorable or unfavorable actions taken at a certain time, affect the update of the $H$-matrix not only for the current state-action pair but also for those met later.
Thus, there is a non trivial interplay between the switching probability observed for the varius states that leads to the observed behavior and that we are unable to unravel completely.

All these considerations find their counterpart in the probability of being at a certain distance from the target during an episode, see Figure~\ref{fig2}c.
Since at the beginning of an episode the agents are introduced uniformly within the region defined by $\sigma/2 < r \leq \tilde{R}$ and the episodes have a limited duration, the radial probability obtained by following the initial policy increases with $r$ in such a region, and has a very slow decay for $r>\tilde{R}$.
On the other hand, the same observable computed for optimized type A agents shows a similar behavior for $r \leq \tilde{R}$ but a quick drop in the external region, suggesting that most of the success achieved by these agents is due to their ability to perform quick tumbles once they are beyond the detection range and prolonged runs in the region where they can detect the distance from the target.
Remarkably, the radial probability displayed by type B agents not only has an even faster decay for $r>\tilde{R}$ but it starts to decrease already for $r\gtrsim 2\sigma$.
Finally, note that for both type A and type B agents the increased tumbling rate in the external region results in a peak of the radial distribution just beyond the detection range.

\begin{figure*}[h!]
\centering
\includegraphics[width=0.9\textwidth]{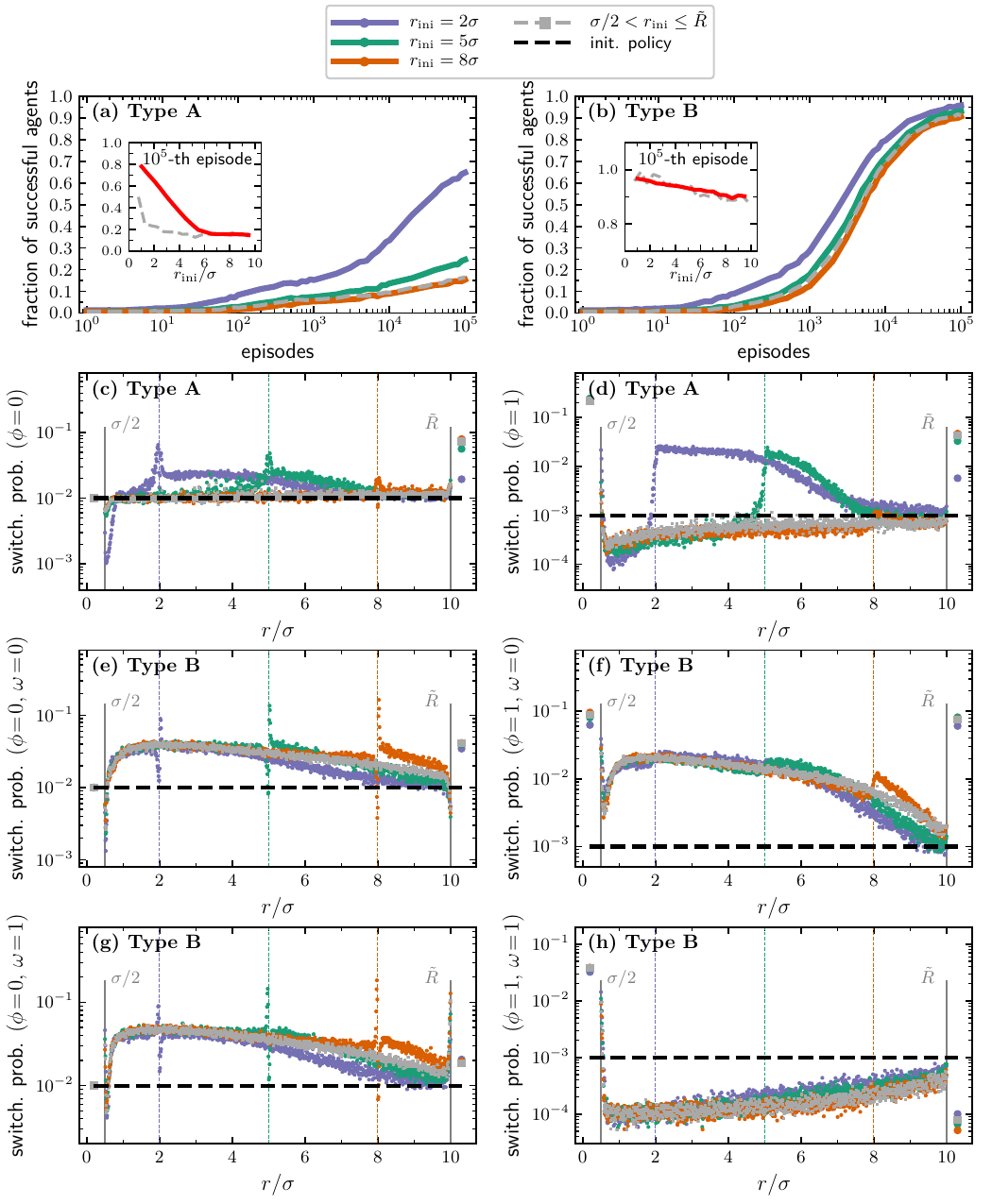}
\caption{
{\bf a)} Fraction of successful Type A agents as a function of the number of episodes, shown for various fixed initial distances from the target. The learning curve for $r_{\rm ini}$ uniformly randomized within the range $\sigma/2 < r \leq \tilde{R}$, previously shown in figure~\ref{fig1}a, is here again reported (grey-dashed line). Inset: Fraction of successful as a function of the initial distance.
{\bf b)} Same as panel (a) but for type B agents.
{\bf c-h)} Switching probabilities at episode $10^5$ for $r_{\rm ini}=2\sigma$ (violet), $r_{\rm ini}=5\sigma$ (green), and $r_{\rm ini}=8\sigma$ (orange).
Across panels (c-h), the corresponding initial policy is reported with a black-dashed horizontal line, the value of $r_{\rm ini}$ is marked with a vertical dashed line of the corresponding color, and the policy learned leaving $r_{\rm ini}$ uncostrained is reported with gray squares.
}\label{fig3}
\end{figure*}

The results reported above show that the detection range $\tilde{R}$ plays an important role in learning a successful strategy. 
An interesting question is whether our learning agents are able to adapt and benefit from a second imposed length scale, as suggested by previous literature~\cite{munozgil2023,Benhamou2015,Ferreira2021}.
We explore such a question by running our learning algorithm in a setup in which the distance from the target at the beginning of each episode, $r_{\rm ini}$, is fixed.
Specifically, we run $18$ different learning processes at fixed $r_{\rm ini}$, varying its value in the range from $\sigma$ to $9.5\sigma$ in steps of $0.5\sigma$.
Figure~\ref{fig3} show the results obtained for three of the different values of the initial distance, namely $r_{\rm ini}=2,\,5,\mbox{ and }8\sigma$, after a learning process lasting $10^5$ episodes.
In line with what already observed in Ref.~\cite{munozgil2023}, we first note that the learning performances displayed by the learning agents improve as the initial distance decreases, see Figure~\ref{fig3}a and b.
This effect is more pronounced for type A agents which are unable to make temporal comparisons.
After $10^5$ episodes, type B agents with a fixed $r_{\rm ini}$ reach a fraction of successful agents as a function of the initial distance which is similar to that obtained with the original protocol having $r_{\rm ini}$ randomly chosen in $(\sigma/2,\tilde{R}]$, see inset of Figure~\ref{fig3}b.
This does not hold for type A agents, where the protocol with a fixed $r_{\rm ini}$ yields improved results at the $10^5$ episode, provided that $r_{\rm ini} \lesssim 6\sigma$, see inset of Figure~\ref{fig3}a.

The policies learned after $10^5$ episodes clearly show the fingerprint of the newly imposed length scale (Figure~\ref{fig3}c-h).
Indeed, both the probability of switching from passive to active and from active to passive for type A agents peak in correspondence of $r_{\rm ini}$ and have increased values (with respect to the initial policy) for $r>r_{\rm ini}$.
This indicates that type A agents tend to tumble more frequently as they move beyond their initial distance from the target.
In particular, the large value of the probability of switching from active to passive for the states with $r>r_{\rm ini}$ and its slow decrease towards the initial policy (see Figure~\ref{fig3}d) point to the fact that the agents aim at shortening the active phases that are bringing them further from the target. 
The signature of the imposed initial distance is less pronounced in the case of type B agents.
Specifically, the probability of switching from active to passive while getting closer to the target ($\omega=1$) is equivalent to that learned in the protocol in which the initial distance is not fixed (Figure~\ref{fig3}h).
With the exception of the case $r_{\rm ini}=8\sigma$, which is probably due to the different learning rates among the various cases, the same holds also when the state of the agents is such that, in the last integration step, it has moved further from the target ($\omega=0$), see Figure~\ref{fig3}f.

Regarding the learned probability for switching from the BP to the ABP phase, we note that, if $\omega=0$ it is quite similar to the one learned when the initial distance is not fixed but it shows a drop just before $r=r_{\rm ini}$ and an increase just after (Figure~\ref{fig3}e).
The opposite holds for $\omega=1$ (Figure~\ref{fig3}g).
This behavior is similar to what is already observed in the proximity of $\tilde{R}$ for the policy learned in the case in which the initial distance is not fixed.
Again, we failed to find a complete explanation for this observed behavior but we note that the agents are able to reproduce it also at the new imposed length scale, corroborating the belief that this behavior plays an important role in optimizing the performances of type B agents.

As a final remark, we note that the above observations on the enhaced learning rates at small imposed initial distances and on the policies suggest that a possible way to speed-up the learning process is to learn a policy at a small $r_{\rm ini}$ and then exploit it as the initial policy during the learning process at a slightly larger value of $r_{\rm ini}$. By iteratively repeating this procedure, one can more efficiently learn a successful policy also at large or randomly selected values of $r_{\rm ini}$ (results not shown -- see Ref.~\cite{Tovazzi2024thesis} for details).

\section{Conclusion}\label{sec4}

When looking for nourishment, bacteria like \textit{E. coli} make no controlled changes of direction.
However, they can sense the concentration of a certain chemoattractant, compare its value to a previous one, and adjust their tumbling rates accordingly.
This results in a run-and-tumble motion characterized by longer runs in the favorable direction, thus enhancing their efficiency in reaching the target.

We tried to better understand how this bacteria behavior arises by exploiting reinforcement learning.
In particular, we considered a learning agent initially performing unbiased run-and-tumble dynamics and applied the Projective Simulation algorithm~\cite{Briegel2012} to enable the agent to develop an efficient stochastic target-search strategy.
The learning agent has been modelled as an intermittent active Brownian particle~\cite{Caraglio2024,Kiechl2024} that can switch from a standard passive Brownian phase to an active Brownian phase and viceversa. 
The transition between these phases occurs with a probability that depends on the agent's internal state and that is tuned during the learning process in order to optimize the target-search performances.
We considered two different types of agents: 
Type A agents can only sense the distance to the target, which in a homogeneous environment is a good proxy for the concentration of ligands released by the target.
Type B agents, beyond sensing the distance to the target, are endowed with a short time memory that allows them to make temporal comparisons similar to those made by bacteria.

Our findings show that both types of agents are able to learn successful target-search policies, with those equipped with temporal comparison abilities achieving significantly better performances.
Furthermore, contrary to what is displayed by type A agents, the efficiency of trained type B agents only slightly depends on the initial distance of the target.
By inspecting the policy learned by type A agents we note that the probability of tumbling increases with the distance to the target and that the active phases should in general be longer than what initially defined based on the \textit{E. coli} behavior in a uniform dilute acqueus medium~\cite{Berg2004}.
The additional information sensed by type B agents results in a clear distinction in the switching probability from the active to the passive phase: Depending on whether the particle is moving towards the target or not, the active phases are respectively prolonged or shortened.
Finally, when the initial distance of the target is fixed during the learning process, the policies learned by our agents clearly display a signature of this additionally imposed length scale.
Consequently, our agents learn to exploit this additional length scale to further improve their efficiency in locating the target, in accordance with what is suggested by previous literature~\cite{munozgil2023,Benhamou2015,Ferreira2021}.

Our work is mainly addressed to investigating how the chemotactic behavior shown by bacteria can be achieved through reinforcement learning.
However, the same framework can also be applied to artificial microswimmers whose activity can be controlled by an external illuminating system~\cite{Muinos-Landin2021}.
While it has already been shown that Janus particles~\cite{Bechinger2016} able to couple their self-propulsion orientation to a chemical gradient can perform chemotaxis~\cite{Popescu2018}, the phase switching mechanism we proposed may represent a valid alternative to endow artificial microswimmers with chemotactic abilities.

Finally, our work can be leveraged to explore more complex and eventually realistic scenarios such as, for instance, bacterial migration through confined spaces and porous media~\cite{Davit2013,Brown2016,Sosa-Hernandez2017} or in front of solid surfaces~\cite{Giacche2010,Guillaume2010}.
Moreover, the randomization of the self-propulsion direction at each phase switch could be suppressed in order to take into account that there are experimental indications that bacteria can also tune the amount of reorientation during the tumbling phases by controlling their duration.
Moreover, the randomization of the self-propulsion direction at each phase switch could be suppressed to reflect experimental evidence suggesting that bacteria can modulate the the amount of reorientation during tumbling by adjusting the duration of this phase~\cite{Turner2000,Saragosti2011}.

\section{Methods}\label{sec5}

To identify effective target-search strategies, we employed the reinforcement learning algorithm Projective Simulation (PS). 
Originally developed for designing autonomous quantum learning agents~\cite{Briegel2012}, PS has demonstrated competitive performance in classical RL problems as well~\cite{Mautner2015,Boyajian2020}.

The key feature of the PS algorithm is the use of a particular memory structure, termed \textit{episodic and compositional memory} (ECM), mathematically represented as a graph of interconnected units called \textit{clips}.
Clips correspond to either perceptual units (state), decision units (actions), or a combination of those.

We modeled the target-search problem as a Markov decision process~\cite{Sutton2018}, where, at each learning step, the agent has a state $s$, performs an action $a$ based on a policy defined by the conditional probabilities $\pi(a|s)$, and receives a reward $\mathcal{R}$ as feedback. 
The ECM structure in this context consists of a layer of states fully connected to a layer of actions.
Each state-action pair $(s,a)$ is associated with a real-valued weight $h(s,a)$, called the $h$-\textit{value}, which defines the policy as:
\begin{align}
	\pi(a|s) = \dfrac{h(s,a)}{\sum_{a' \in \mathcal{A}}h(s,a')} \; ,
\end{align}
where $\mathcal{A}$ is the set of all possible actions.
Additionally, a non-negative \textit{glow value} $g(s,a)$ tracks the frequency and recency of visits to specific state-action pairs, influencing the policy updates to optimize the total expected reward.
This glow-based updating mechanism makes PS particularly well-suited for our target-search problem: 
Since a large number of iterations of the motion equations~(\ref{eom1}-\ref{eom3}) are required before finding the target and receiving a reward, the reward signal is sparse and weakly correlated with individual state-action pairs. 
Methods capable of processing sequences of state-action pairs, such as PS, are thus more effective than traditional action-value algorithms like Q-learning or SARSA~\cite{Sutton2018}, which failed to produce successful policies in a similar setup~\cite{Caraglio2024}.

In our model (see dedicated section), the action $a$ is binary: $a=1$ triggers a phase switch, while $a=0$ maintains the current phase.
Applying the PS framework with a learning iteration at each integration of the equations of motion, Eqs.~(\ref{eom1}-\ref{eom3}), each learning step proceeds as follows:
\begin{itemize}
	\item The agent determines the phase-switching probability $p_t$ for the current state $s_t$ as:
	\begin{align}
		p_t := \pi(a=1|s_t) = \dfrac{h(s_t,1)}{h(s_t,0)+h(s_t,1)} \; ,
	\end{align}
    and selects the next phase $\phi_{t+\Delta t}$ accordingly;
    \item The glow matrix $G$ is damped according to 
    \begin{align}
    	G \leftarrow (1-\eta)G \; ,
    \end{align}
    where $\eta$ is called the glow parameter and determines how much a delayed reward should be discounted;
    \item A unit is added to the glow value of the current state-action pair 
    \begin{align}
    	g(s_t,a_t) \leftarrow g(s_t,a_t)+1 \; ;
    \end{align}
    \item The particle's position and sef-propulsion direction at time $t+\Delta t$ are computed according to Eqs.~(\ref{eom2}-\ref{eom3});
    \item The $h$-value matrix $H$ is updated as 
    \begin{align}
    	H \leftarrow (1-\gamma) H +\gamma H_0 + \mathcal{R}G\; ,
    \end{align}
    where $\mathcal{R}$ is the reward (being 1 if $r_{t + \Delta t} \leq \sigma/2$ and 0 otherwise), and the damping parameter $\gamma$ controls the rate of return to an initial matrix $H_0$.
\end{itemize}
The glow and damping parameters are treated as hyperparameters and tuned for optimal learning performance.
Along the manuscript we used $(\gamma,\eta)=(10^{-6},10^{-4})$ both for type A amd type B agents.

To ensure a finite number of states, the distance to the target $r$ is binned using a bin width equal to $0.01\sigma$ in the interval $\sigma/2 < r \leq \tilde{R}$ and two more bins are defined for $r \leq \sigma/2$ and $r>\tilde{R}$ respectively.

The initial policy assigns probabilities of phase switching as $10^{-2}$ and $10^{-3}$ for passive and active phases, respectively, achieved by setting the $h_0$-values accordingly.

The time step is $\Delta t=10^{-4}\tau$.

Similarly to what has been done in a previous paper~\cite{Caraglio2024}, we optimize computational efficiency by exploiting the fact that the reward is different from zero only when the target is found and that finding the target terminates the current episode.
We then update the $H$-matrix only at the end of an episode according to
\begin{align}
	H \leftarrow (1-\gamma)^{n_{\rm ep}} H + \gamma \left[ \sum_{i=0}^{n_{\rm ep}-1} (1-\gamma)^i \right] H_0 + \mathcal{R} G \; ,
\end{align}
where $n_{\rm ep}$ is the number of learning steps within the given episode.
During intermediate steps within the episode $n := t/\Delta t \leq n_{\rm ep}$, the phase-switching probability $p_t$ is computed according to
\begin{align}
	p_t  = \dfrac{\tilde{h}(s_t,1)}{\tilde{h}(s_t,0)+\tilde{h}(s_t,1)} \; ,
\end{align}
using the temporarily updated $h$-values:
\begin{align}
	\tilde{h}(s_t,a_t) := & (1-\gamma)^{n-1} h(s_t,a_t)  \nonumber \\
	  & \quad + \gamma \left[ \sum_{i=0}^{n-2} (1-\gamma)^i \right] h_0(s_t,a_t) \; .
\end{align}
It is also computationally extremely more efficient to save, for each state-action pair $(s,a)$, the time steps $n_1^{(s,a)},n_2^{(s,a)},\ldots,n_{M^{(s,a)}}^{(s,a)}$ at which they are visited and then also update the glow matrix only at the end of the episode according to
\begin{align}
	g(s,a)  \leftarrow \left\lbrace 
	\begin{array}{ll}
		(1-\eta)^{n_{\rm ep}} g(s,a) & \; \mbox{if } M=0 \, , \\
		& \\
		(1-\eta)^{n_{\rm ep}-n_1}  & \\
		\; \times \! \left[ (1-\eta)^{n_1} g(s,a) \!+\! 1\right]  & \; \mbox{if } M=1 \, , \\
		& \\
		(1-\eta)^{n_{\rm ep}-n_2} \left\lbrace  (1-\eta)^{n_2} \right.  & \\
		\; \times \!  \left.  \left[ (1-\eta)^{n_1} g(s,a) \!+\! 1\right] \!+\! 1 \right\rbrace  & \; \mbox{if } M=2 \, , \\
		& \\
		\qquad \ldots & \\
	\end{array}
	\right. 
\end{align}
where we dropped the superscripts $(s,a)$ for the sake of compactness.
All elements of the $G$ matrix are initialized to zero every 20 episodes.

\end{document}